\documentclass[twocolumn,twoside]{IEEEtran}
\ifCLASSINFOpdf
  
\else
 
\fi

\ifCLASSOPTIONcompsoc
    \usepackage[caption=false, font=normalsize, labelfont=sf, textfont=sf]{subfig}
\else
    \usepackage[caption=false, font=normalsize]{subfig}
\fi
\usepackage{lipsum}%

\usepackage{balance}
\usepackage{multicol}   
\usepackage{cite}
\usepackage{gensymb}
\usepackage{multirow}
\usepackage{graphics}  
\usepackage{epsfig} 
\usepackage{graphicx}
\usepackage{epstopdf}
\usepackage{textcomp}
\usepackage{amsmath}
\usepackage{mathtools}
\usepackage{filecontents}
\usepackage{lipsum,color}
\usepackage{amssymb}
\usepackage{float}
\usepackage{stfloats}
\usepackage{blindtext}
\usepackage{hyperref}
\usepackage{eso-pic}

\makeatletter
\def\ps@IEEEtitlepagestyle{%
	\def\@oddfoot{\mycopyrightnotice}%
	\def\@evenfoot{}%
}
\def\mycopyrightnotice{%
	{\hfill \footnotesize 978-1-7281-8012-0/20/\$31.00 \copyright 2020 IEEE\hfill}
}
\makeatother

\begin{document}
	 \AddToShipoutPictureFG*{
		\AtPageUpperLeft{\put(0,-10){\makebox[\paperwidth][l]{2020 10th International                 Symposium on Telecommunications (IST'2020)}}}
	}
\title{Statistical Channel Modeling for Long-Range Ground-to-Air FSO Links}

\author{Hossein~Safi,~and~Akbar~Dargahi
\thanks{Hossein Safi is with the Department of Electrical Engineering, Shahid Beheshti University G. C., 1983969411, Tehran, Iran, (e-mail: h\_safi@sbu.ac.ir). Akbar Dargahi is with the Department of Electrical Engineering, Shahid Beheshti University G. C., 1983969411, Tehran, Iran (e-mail: a-dargahi@sbu.ac.ir) }}


\maketitle

\begin{abstract}
To provide high data rate aerial links for 5G and beyond wireless networks, the integration of free-space optical (FSO) communications and aerial platforms has been recently suggested as a practical solution. To fully reap the benefit of aerial-based FSO systems, in this paper, an analytical channel model for a long-range ground-to-air FSO link under the assumption of plane wave optical beam profile at the receiver is derived. Particularly, the  model includes the combined effects of transmitter divergence angle, random wobbling of the receiver, jitter due to beam wander, attenuation loss, and atmospheric turbulence. Furthermore, a closed-form expression for the outage probability of the considered link is derived which makes it possible to evaluate the performance of such systems. Numerical results are then provided to corroborate the accuracy of the proposed analytical expressions, and to prove the superiority of the proposed channel model over the previous models in long-range aerial FSO links.
\end{abstract}
\IEEEpeerreviewmaketitle
\section{Introduction}
More recently, the idea of integrating free space optical
(FSO) communications with aerial platforms, i.e., balloons, airships, drones, etc. has been the subject of many studies in academia and industry to establish high capacity and easy-to-deploy low power consumption links \cite{alzenad2018fso,fawaz2018uav}. Such systems seem capable of carrying out multiple tasks, especially they can be utilized as aerial
stations to enable a ubiquitous connectivity for
the next generation of wireless networks \cite{safi2019spatial}.

However, to fully obtain the benefits of airborne FSO links, the communication channel should be explicitly  characterized. There has recently been an increasing amount of literature on this topic \cite{li2018investigation, dabiri2018channel,dabiri2019tractable,najafi2018statistical}. For instance, a theoretical model for a link between an unmanned aerial vehicle (UAV) and a satellite  is developed in \cite{li2018investigation} when the effects of
Doppler frequency shift, pointing error, and atmospheric turbulence are taken into account. The work in \cite{dabiri2018channel} proposes a complex channel model for a short link airborne FSO system with which the effect of transceivers' wobbling (also termed vibrations) on the link performance is investigated in detail. Meanwhile, for the considered system model in \cite{dabiri2018channel}, the authors further develop their results and derive a more tractable channel model in \cite{dabiri2019tractable}. Najafi \textit{et al.}, propose a statistical model for the geometric loss of a UAV-based FSO link in \cite{najafi2018statistical}, and also analyze the performance of such link in terms of outage probability and ergodic rate under different turbulence conditions in \cite{najafi2019statistical}.

Nevertheless, these prior works mainly concern short-range aerial FSO links, and thus, their proposed pointing error models have been developed under the assumption of the Gaussian profile for the received optical beam. Since plane wave and spherical
wave models are more accurate optical wave models for characterizing the beam profile in long-range FSO links \cite{saleh2019fundamentals,laserbook}, these proposed channel models, however, may not be directly applicable for long-range FSO links. In addition, prior studies \cite{dabiri2018channel,dabiri2019tractable,najafi2018statistical,najafi2019statistical,khankalantary2020ber1} have not been able to account for all aspects of link parameters, e.g, the effect of beam wander on the link performance has not been investigated.

To fill this gap, in this paper, we consider a long-range ground-to-air FSO link\footnote{For example, a link to an aerial platform above the stratosphere layer (more than 20 km above the ground level)} and derive a new model for pointing error under the assumption of plane wave for the optical beam profile at the receiver. In addition to the pointing error that includes the effects of receiver's wobbling and beam wander, we take the effects of atmospheric turbulence and attenuation loss into account and derive an analytical channel model for the considered system.  Moreover, we derive a closed-form expression for the outage probability
of the considered link under strong atmospheric turbulence conditions. We report simulation results to corroborate the accuracy
of the derived analytical expressions. We then show that, unlike the proposed model in \cite{dabiri2018channel}, our proposed analytical-based model can achieve a perfect match with the exact simulation-based results. The developed results of this paper can thus be applied for studying and designing long-range FSO links without resorting
to time-consuming simulations.

The rest of this paper is organized as follows. In Section \ref{sysmodel}, we present the system model and the main assumptions considered in this paper. Next, in Section \ref{chanmodel}, we provide analytical formulations for the channel model and the outage probability of the considered ground-to-air FSO link. We presents numerical results
to verify our analytical expressions and demonstrate
the need of an accurate channel model for our considered setup in Section \ref{num}. Finally, Section \ref{con} concludes the paper.

\section{System Model}
\label{sysmodel}

The positions of the receiver and transmitter are determined by $H_{Tx} = (0,0,0)$ and $H_{Rx} = (l_x,l_y,Z+l_z)$, respectively. It is assumed that the receiver node is hovering at a fixed altitude, and thus, there is no Doppler spreading effect. The independent random variables (RVs) $l_x$, $l_y$, and $l_z$, which are zero-mean Gaussian distributed having  variance $\sigma^2_l$, denote the random deviations due to the receiver wobbling along the axes of coordinates \cite{dabiri2018channel,UAVmmwave2020}. In comparison with the link length $Z$, the variation of $l_z$ is negligible, thus the effect of pointing error due to receiver wobbling around the $z$-axis can be reasonably neglected. However, for long-range FSO links, the effect of optical beam wander induced pointing errors should be taken into account \cite{laserbook}. Accordingly, at the receiver aperture, the displacements of the optical beam because of beam wander effects along the $x$ and $y$ coordinates, denoted by $b_x$ and $b_y$, are RVs with Gaussian
distribution having zero-mean and variance \cite{li2018investigation}
\begin{align}
\label{variance_beam_wander}
\sigma_{b}^2 = 2.07
\int_{h_0}^{H} C_n^2\left( h \right)\left(Z-h\right)^2w_h^{-\frac{1}{3}}dh,
\end{align}
where $w_h = \frac{h\theta_\textrm{div}}{2}$, and $C_n^2\left( h \right)$ is the refractive-index structure parameter as a function of the altitude $h$.

 Employing intensity modulation and direct detection (IM/DD) techniques,
at the $k$th symbol interval, the output photo-current can be written as 
\begin{align}
	\label{sg1}
	r[k] = \eta\, h\, s[k]  +  n[k],
\end{align}
where $\eta$, $h$, $s[k]$, and $n[k]$ denote photo-detector (PD) responsivity, the channel gain, the transmitted symbol with average optical power $P_t$, and  the Gaussian noise having zero mean and variance $\sigma_{n}^2$, respectively. 
Moreover, in an FSO system, the instantaneous electrical signal-to-noise ratio (SNR) can be written as
 \begin{align}
 \label{SNR}
 \gamma = \dfrac{\eta^2P_t^2h^2}{\sigma_n^2}.
 \end{align}
In practical FSO systems the coherence time of the channel seems quite long relative to the bit time \cite{safi2019adaptive}. Thus, for such systems with slow fading property, outage probability (i.e., the probability with which the instantaneous SNR falls bellow a certain threshold $\gamma_\textrm{th}$) is an appropriate parameter for evaluating the link performance \cite{pointing2007}. Thereby, the outage probability can be written as
\begin{align}
\label{outage}
\mathbb{P}_{\textrm{out}} = \int_{0}^{\gamma_\textrm{th}} f_{\gamma}(\gamma) d\gamma = \int_{0}^{\frac{\sqrt{\gamma_\textrm{th}\sigma_n^2}}{\eta P_t}} f_{h}(h) dh,
\end{align}
where $f_{\gamma}(\gamma)$ and $f_h(h)$ denote the probability density function (PDF) of $\gamma$ and $h$, respectively.
\section{ Channel Modeling}
\label{chanmodel}
The channel gain  $h$ can be written as
\begin{align}
	\label{hj5}
	h = {h_{a} h_{t} h_{p}},
\end{align}
where the parameters $h_{a}$, $h_{t}$, and $h_{p}$, denote the attenuation loss, the atmospheric turbulence,
and the effective pointing error induced geometrical loss, respectively. In the sequel, we discuss about these channel parameters in more detail.
\subsubsection{Attenuation loss and Atmospheric Turbulence}
The attenuation loss can be modeled by the  Beers-Lambert law as $h_{a}=\exp\left(-Z\xi\right)$, where $\xi$ denotes the attenuation coefficient which is a function of the visibility \cite{al2004fog}.
The optical turbulence can be modeled as Gamma-Gamma distribution which is the dominant
statistical fading model for FSO links because of its excellent agreement with measurement
data for weak to strong turbulence conditions\cite{laserbook}. Thus, the PDF of $h_{t}$ can be written as 
\begin{eqnarray}
\label{fg1g}
f_{\rm {GG}}(h_{t})=  \frac{2(\alpha\beta)^{\frac{\alpha+\beta}{2}}}{\Gamma(\alpha)\Gamma(\beta)} 
h_{t}^{\frac{\alpha+\beta}{2}-1}   
K_{\alpha-\beta}(2\sqrt{\alpha\beta h_{t}}) 
\end{eqnarray}
where $ \Gamma (\cdot) $ denotes the Gamma function, and $ K_{\delta}(\cdot) $ represents the modified Bessel function of the second kind of order $\delta$. Furthermore, $ \alpha $ and $ \beta $ are the parameters related to the number of large-scale and small-scale eddies, respectively.
\begin{figure}
	\begin{center}
		\includegraphics[width=3.75 in ]{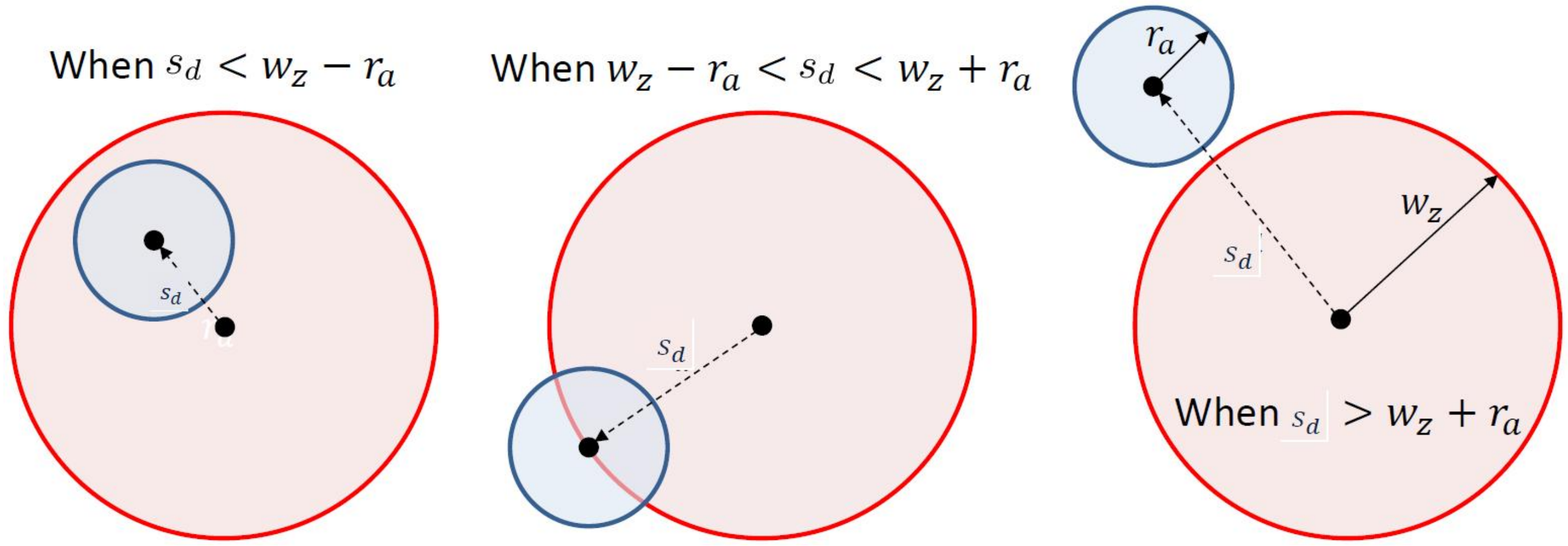}
		\caption{ Beam footprint (red circle) at the receiver aperture (blue circle). The center of the received beam is deviated from the center of receiver lens due to beam wander effects and the random wobbling of hovering receiver node.}
		\label{pointing error}
	\end{center}
\end{figure}
\subsubsection{Effective Pointing Error}




The receiver aperture with area $A_r = \pi r_a^2$ has been located at a large distance $Z$ from the transmitter, i.e., $Z\gg r_a$. Thus, at the receiver the phase difference over the optical wavefront can be reasonably omitted 
and the paraboloidal phase factor of the optical beam approaches zero.
According to \cite[(1.2.5) and (1.2.6)]{karp2013optical}, 
when the optical transmitter radiates all its power through the two-dimensional solid angle $\Omega_s$, the field intensity at distance $Z$ is 
\begin{align}
\label{int1}
I(Z) = \frac{P_t}{\Omega_s Z^2}.
\end{align}
Considering $\theta_\textrm{div}$  as the divergence angle of the transmitted beam,  the radius of received optical beam footprint at the receiver aperture can be approximated as  $w_z =\dfrac{\theta_\textrm{div} Z}{2}$, we also have $\Omega_s = 2\pi\left( 1-\cos\left(\frac{\theta_\textrm{div}}{2}\right) \right)$. 
Accordingly,  when incident optical beam is orthogonal to the PD plane, the amount of collected power by the receiver with a collecting aperture area of $A_r = \pi r_a^2$ is 
\begin{align}
\label{cb1}
P_r = \frac{ r_a^2 P_t }{2\left( 1-\cos(\frac{\theta_\textrm{div}}{2})\right)Z^2}.
\end{align}

Let $s_d$ represents the separation distance between the center of optical beam footprint and the center of the receiver aperture. As a result, the effective pointing error as a function of $s_d$ is formulated as
\begin{align}
\label{zxz}
h_{p}\left( s_d \right) =& \frac{ r_a^2 \cos \left(\theta_d\right) u(w_z-r_a-s_d)}{2\left( 1-\cos\left(\frac{\theta_{div}}{2}\right)\right)Z^2} \\ & +\frac{ A_i(s_d,\theta_d)  \left(u(w_z-r_a-s_d) - u(w_z+r_a-s_d)\right)}{2\pi\left( 1-\cos\left(\frac{\theta_{div}}{2}\right)\right)Z^2}, \nonumber
\end{align}
where $u(\cdot)$ is the well-known step function, and  $A_i\left(s_d\right)$ is the overlapping area between the receiver lens and received beam when $ w_z-r_a \leq s_d\leq w_z+r_a$ (i.e., circle-circle intersection) and it can be obtained as 
\begin{align}
\label{aa}
&A_i\left(s_d\right) = 
-\frac{1}{2}\sqrt{\left(-s_d+r_a+w_z\right)\left(s_d+r_a-w_z\right)} \\
&\times\sqrt{\left(s_d-r_a+w_z\right)\left(s_d+r_a+w_z\right)} \nonumber \\
&+r_a^2 \cos^{-1}\left( \frac{s_d^2+r_a^2-w_z^2}{2s_d r_a} \right) 
+w_z^2 \cos^{-1}\left( \frac{s_d^2+w_z^2-r_a^2}{2s_d w_z} \right). \nonumber
\end{align} 
To find the PDF of the effective pointing error, $f_{h_{p}(h_{p})}$, we resort to the following theorem.

{\bf Theorem 1.} 
\emph{The PDF of $h_{p}$ is analytically derived as \eqref{fag}, where $\delta(\cdot)$ is the Dirac delta function and  $\Pi(\cdot)$ is  the gate function defined as
\begin{equation}
\label{vcv1}
\Pi(x) = \left\{
\begin{array}{rl}
1 &~ \text{if }~ x < 1\\
0 &~ \text{if }~ x > 1
\end{array} \right..
\end{equation}
}\emph{Moreover, in \eqref{fag},
$\mathcal{C}_3 = \mathcal{C}_2 \left(\pi w_z r_a - r_a^2+w_z^2 \right)$,
$\mathcal{C}_2 = \left( \frac{\mathcal{C}_1}{w_z r_a} \right)$,
$\mathcal{C}_1=\frac{ r_a^2     }     {4\pi\left( 1-\cos\left(\frac{\theta_{div}}{2}\right)\right)Z^2}$, 
$h_{pg1}=\mathcal{C}_1  \left(\pi - \frac{2r_a^2+2r_a w_z}{w_z r_a} \right)$, and
$h_{pg2}=\mathcal{C}_1    \left(\pi - \frac{2r_a^2-2r_a w_z}{w_z r_a} \right)$. }
\begin{figure*}
	\normalsize
	\begin{align}
	\label{fag}
	f_{h_{p}}(h_{p}) \simeq& \left[1-\exp\left(-\frac{(w_z-r_a)^2}{2\sigma_r^2}\right)\right] 
	\times \delta\left(h_{p}-2\pi\mathcal{C}_1\right)
	+ \exp\left(-\frac{(w_z+r_a)^2}{2\sigma_r^2}\right)  \times \delta\left(h_{p}\right) \nonumber \\
	&+ \frac{1}{2 \mathcal{C}_2 \sigma_r^2}\exp\left( \frac{ h_{p}-\mathcal{C}_3}{2 \mathcal{C}_2 \sigma_r^2}  \right)
	\times\left[\Pi\left(\frac{h_{p}}{h_{pg2}}\right) -\Pi\left(\frac{h_{p}}{h_{pg1}}\right) \right].
	\end{align} 
	\hrulefill
	\vspace*{4pt}
\end{figure*}
\begin{IEEEproof}
We have $s_d=\sqrt{(l_x + b_x)^2+(l_y + b_y)^2} $. Since $l_x$ and $l_y$ Gaussian RVs having zero mean and variance $\sigma_l^2$, the RV $s_d$ can be modeled by a
Rayleigh distribution shown as
\begin{align}
\label{zw}
f_{s_d}(s_d)=\frac{s_d}{\sigma_r^2}\exp\left(-\frac{s_d^2}{2\sigma_r^2}\right),~~~~s_d\geq 0
\end{align} 
where $\sigma_r^2 = \sigma_l^2 + \sigma_b^2$. 
From \eqref{zxz}, the PDF of $h_{p}$ can be formulated as
\begin{align}
\label{xv}
f_{h_{p}}(h_{p}) = f_{1h_{p}}(h_{p})\! + \!f_{2h_{p}}(h_{p})\! + \!f_{3h_{p}}(h_{p}),
\end{align}
where $f_{1h_{p}}(h_{p})$, $f_{2h_{p}}(h_{p})$, and $f_{3h_{p}}(h_{p})$ are the distribution of $h_{p}$ when $(s_d\leq w_z-r_a)$, $( w_z-r_a \leq s_d\leq w_z+r_a)$, and $(s_d\geq w_z+r_a)$, respectively. 
From \eqref{zxz} and \eqref{xv},  $f_{1h_{p}}(h_{p})$ and $f_{3h_{p}}(h_{p})$ can be obtained respectively as
\begin{align}
\label{xb}
f_{1h_{p}}(h_{p}) =&~ \left(1-\exp\left(-\frac{(w_z-r_a)^2}{2\sigma_r^2}\right)\right) \nonumber \\
&\times \delta\left(h_{p}-\frac{ r_a^2  }{2\left( 1-\cos\left(\frac{\theta_{div}}{2}\right)\right)Z^2}\right),
\end{align}
and
\begin{align}
\label{xb3}
f_{3h_{p}}(h_{p}) =\exp\left(-\frac{(w_z+r_a)^2}{2\sigma_r^2}\right) 
\times \delta\left(h_{p}\right).
\end{align}

Calculating $f_{2h_{p}}(h_{p})$ by substituting $A_i\left(s_d\right)$ from \eqref{aa} is mathematically cumbersome. In practical situations, $r_a$ is on the order of cm (2-10 cm), and, as shown in the numerical result section, to reduce the effect of geometrical spreading loss via adjusting the divergence angle of transmitted beam, $w_z$ must be set on the order of several meters.
Under these conditions, eq. \eqref{aa} can be approximated as
\begin{align}
\label{ks}
A_i\left(s_d\right) \simeq r_a^2 \left(\frac{\pi}{2} - \frac{s_d^2+r_a^2-w_z^2}{2w_z r_a} \right)   .
\end{align}
Substituting \eqref{ks} in \eqref{zxz}, for $w_z-r_a \leq s_d\leq w_z+r_a$, $h_{p}$ can be approximated as
\begin{align}
\label{yu}
h_{p} \simeq&~ \frac{ r_a^2 \left(\frac{\pi}{2} - \frac{s_d^2+r_a^2-w_z^2}{2w_z r_a} \right)    }{2\pi\left( 1-\cos\left(\frac{\theta_{div}}{2}\right)\right)Z^2}
.
\end{align}
Now, from \eqref{yu} and \eqref{zw} and after some manipulations, ${\rm for}~h_{pg1}<h_{p}<h_{pg2}$ we have
\begin{align}
\label{uy}
&f_{2h_{p}|\theta_d}(h_{p}) \simeq \frac{1}{2 \mathcal{C}_2 \sigma_r^2}\exp\left( \frac{ h_{p}-\mathcal{C}_3}{2 \mathcal{C}_2 \sigma_r^2}  \right),
\end{align}
where
$\mathcal{C}_3 = \mathcal{C}_2 \left(\pi w_z r_a - r_a^2+w_z^2 \right)$,
$\mathcal{C}_2 = \left( \frac{\mathcal{C}_1}{w_z r_a} \right)$,
$\mathcal{C}_1=\frac{ r_a^2    }     {4\pi\left( 1-\cos\left(\frac{\theta_{div}}{2}\right)\right)Z^2}$. Subsequently, by substituting \eqref{xb}, \eqref{xb3} and \eqref{uy} in \eqref{xv}, $f_{h_{p}}(h_{p})$ is obtained as \eqref{fag} where  
$h_{pg1}=\mathcal{C}_1  \left(\pi - \frac{2r_a^2+2r_a w_z}{w_z r_a} \right)$, and
$h_{pg2}=\mathcal{C}_1    \left(\pi - \frac{2r_a^2-2r_a w_z}{w_z r_a} \right)$. 

\end{IEEEproof}

Finally, by taking the effects of atmospheric turbulence and attenuation loss into account, the PDF of $h=h_{a} h_{t} h_{p}$ can be obtained as
\begin{align}
\label{by}
f_{h}(h) =& \int  f_{h|h_{t}}(h)   f_{h_{t}}(h_{t}) dh_{t}  \\
= & \int  \frac{1}{h_{a} h_{t}}f_{h_{p}}\left(\dfrac{h}{h_{a} h_{t}}\right)   f_{h_{t}}(h_{t}) dh_{t}.\nonumber
\end{align}
Substituting \eqref{fg1g} and \eqref{fag} in \eqref{by} and after some algebra, the analytical expressions of $f_{h}(h)$ is obtained in \eqref{fag4}. 
\begin{figure*}
	\normalsize
	\begin{align}
	\label{fag4}
	f_{h}(h) =&    
	~e^{-\frac{(w_z+r_a)^2}{2\sigma_r^2} } \delta(h) 
		+ \frac{2}{\Gamma(\alpha)\Gamma(\beta)}
		\left(  \frac{\alpha\beta}{2\mathcal{C}_1h_{a}}   \right)^{\frac{\alpha+\beta}{2}}     \left[1-e^{-\frac{(w_z-r_a)^2}{2\sigma_r^2}}\right] 
		h^{\frac{\alpha+\beta}{2}-1}   
		k_{\alpha-\beta}\left(2\sqrt{\alpha\beta \left(\frac{h}{2\mathcal{C}_1h_{a}}\right)}\right)
    \nonumber \\
	&+\frac{(\alpha\beta)^{\frac{\alpha+\beta}{2}}}{h_{a}  \mathcal{C}_2 \sigma_r^2 \Gamma(\alpha)\Gamma(\beta)}     
		\int_{\frac{h}{h_{a} h_{pg2}}}^{\frac{h}{h_{a} h_{pg1}}} 
		\exp\left( \frac{ \frac{h}{h_{a} h_{t}}-\mathcal{C}_3}{2 \mathcal{C}_2 \sigma_r^2}  \right)  
		h_{t}^{\frac{\alpha+\beta}{2}-2}   
		k_{\alpha-\beta}(2\sqrt{\alpha\beta h_{t}}) dh_{t}.
	\end{align} 
	\hrulefill
	\vspace*{4pt}
\end{figure*}

\begin{figure}
	\centering
	\subfloat[] {\includegraphics[width=3.5 in]{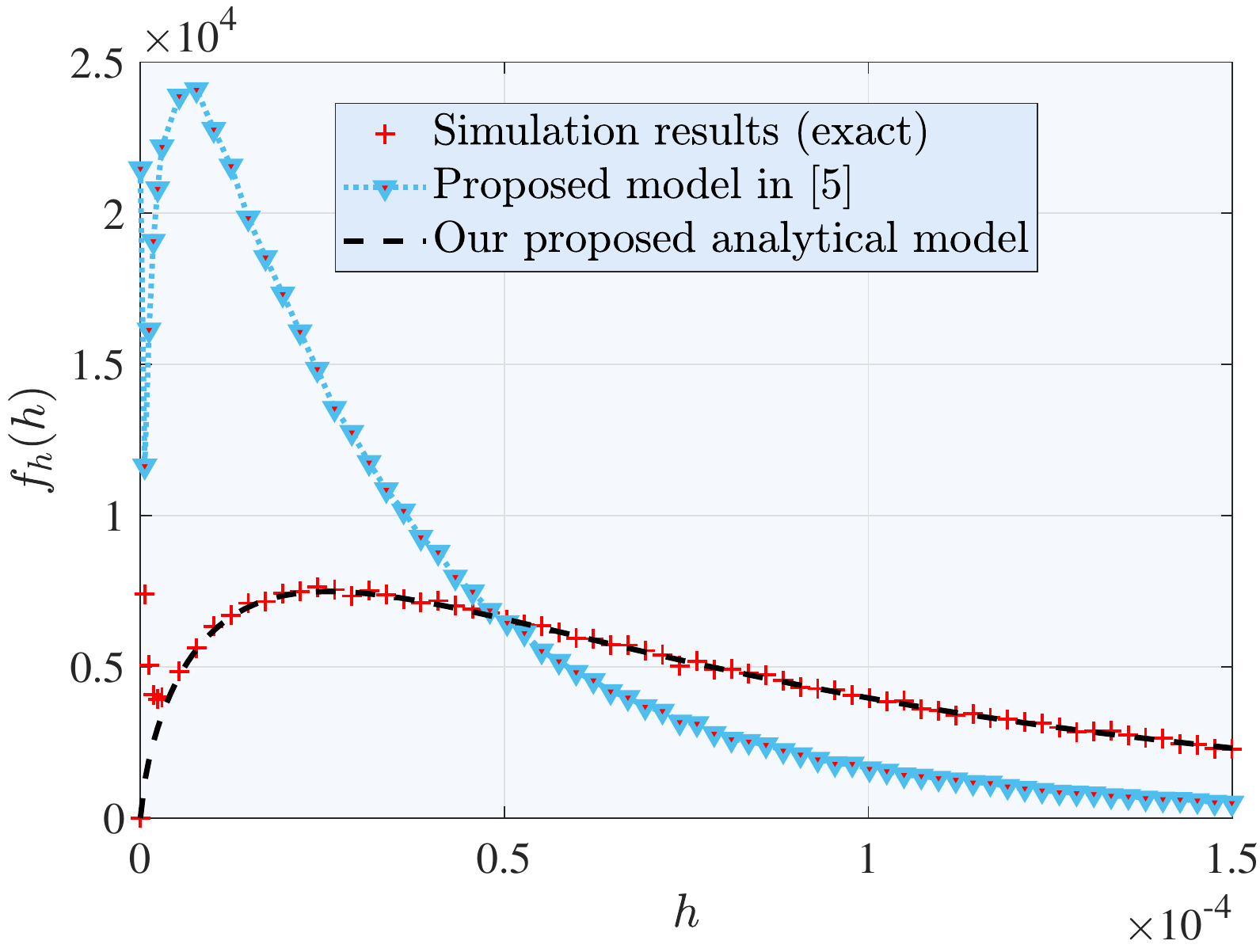}
		\label{6_1}
	}
	\hfill
	\subfloat[] {\includegraphics[width=3.5 in]{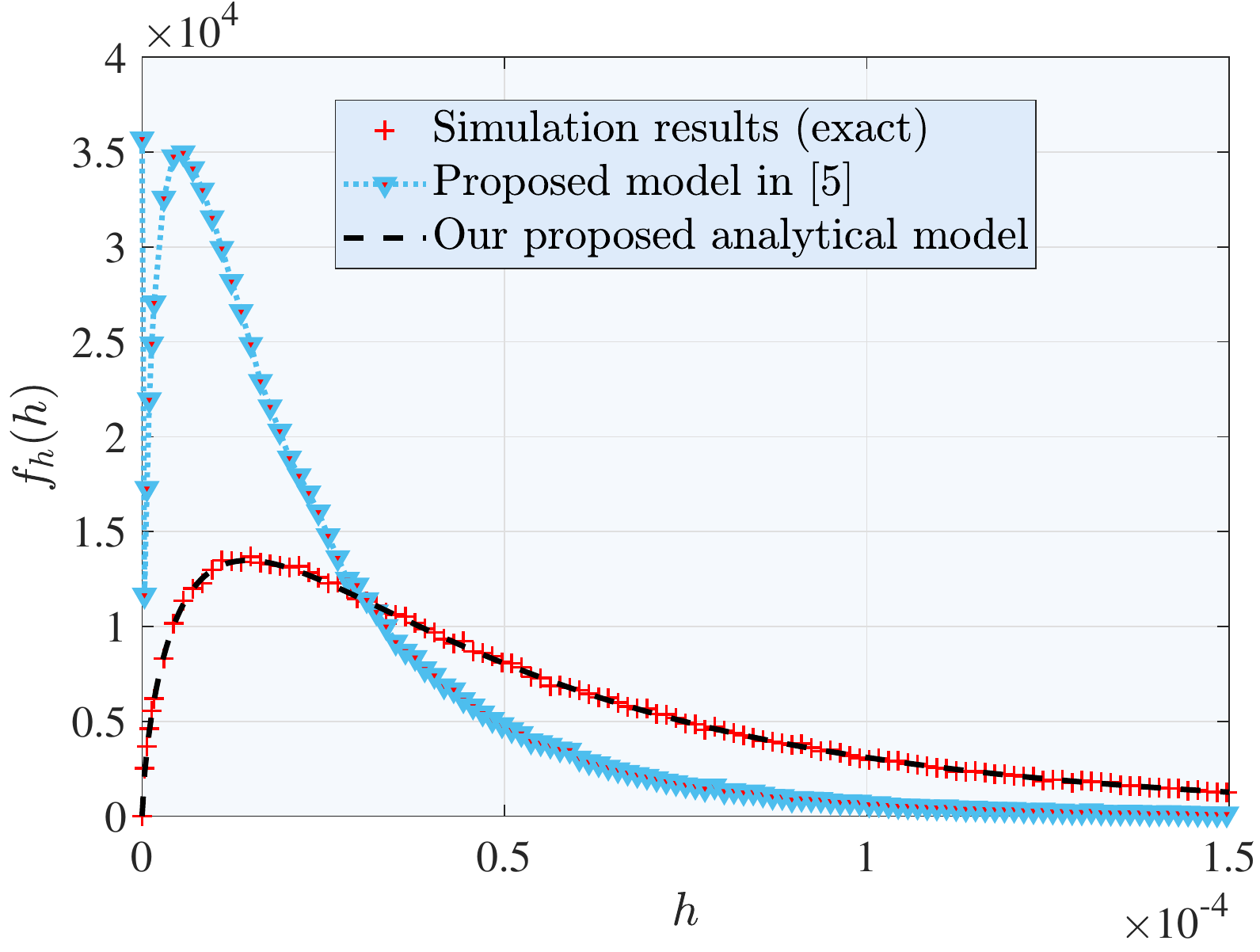}
		\label{6_2}
	}
	\caption{Channel distribution $f_h(h>0)$, for $r_a=5$ cm, $Z =$ 20 km, and a) $\theta_\textrm{div} = 0.1 ~\textrm{mrad}$ \& $w_z/r_a = 20$ and,  b) $\theta_\textrm{div} = 0.2~ \textrm{mrad}$ \& $w_z/r_a = 40$.}
	\label{df2}
\end{figure}

\begin{figure}[t]
	\begin{center}
		\includegraphics[width=3.4 in]{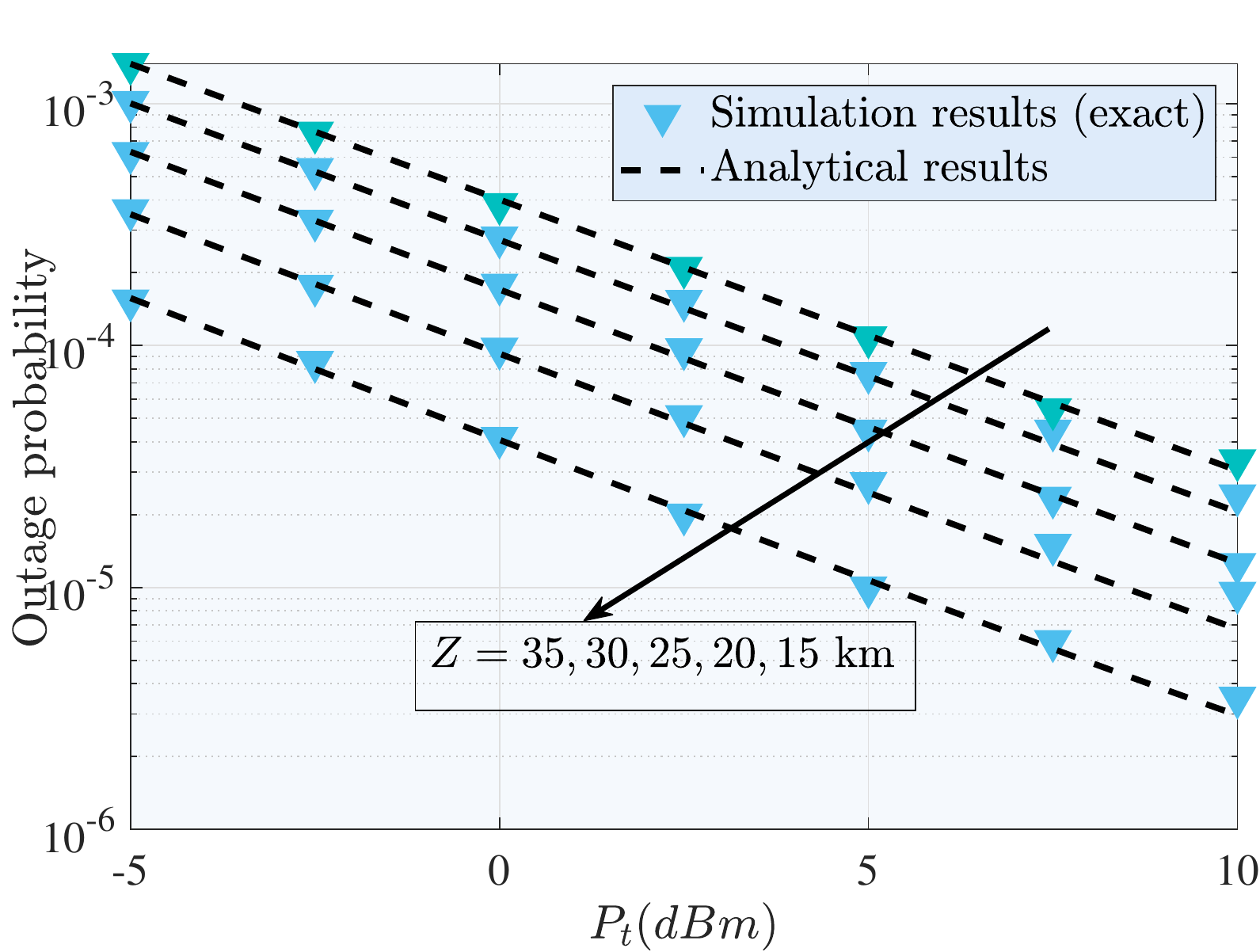}
		\caption{Outage probability versus $P_t$ for $r_a = 10 \textrm{~cm}$, $\theta_\textrm{div} = 0.2 $ mrad, and  different values of link length $Z$.}
		\label{outage_div}
	\end{center}
\end{figure}

Finally, to derive a mathematically tractable closed-form expression for the outage probability of the considered link, we employ the expansion of the modified Bessel function. Accordingly, from \cite{wolfram}, the
modified Bessel function $K_\delta(x)$ can be expanded as  
\begin{align}
\label{besel}
K_\delta(x) \!=\! \dfrac{\pi}{2\sin(\pi \delta)}\! \sum_{n=0}^\infty\! 
\left[\!\dfrac{(x/2)^{2n-\delta} }{\Gamma(n-\delta+1)n!}\!  -  \!  \dfrac{(x/2)^{2n+\delta} }{\Gamma(n+\delta+1)n!} \!\right]\!.
\end{align}
According to \eqref{fag4} and \eqref{besel}, and after some manipulations, the outage probability can be written as
\begin{align}
\label{out3}
\mathbb{P}_{\textrm{out}} =&  \sum_{n=0}^\infty \dfrac{{\mathcal{C}_5}}{2\sin\big(\pi(\alpha-\beta)\big)} \left( \dfrac{h_\textrm{th}^{n+\beta}}{\mathcal{C}_{n1}}  - \dfrac{h_\textrm{th}^{n+\alpha}}{\mathcal{C}_{n2}} \right),
\end{align}
where $\mathcal{C}_{n1} = (n+\beta)\Gamma(n-\alpha+\beta+1)n!$, $\mathcal{C}_{n2} = (n+\alpha)\Gamma(n+\alpha-\beta+1)n!$, and $
\mathcal{C}_5 = \frac{2\pi}{\Gamma(\alpha)\Gamma(\beta)}
\mathcal{C}_4^{\frac{\alpha+\beta+1}{2}}     \left[1-e^{-\frac{(w_z-r_a)^2}{2\sigma_r^2}}\right]$.
\section{Simulation Results and Analysis}
\label{num}
We verify the accuracy of the provided analytical expressions by performing Monte Carlo simulations with over $50 \times 10^6$ independent runs. For simulations, we consider typical system parameters  as follows. The optical wavelength $\lambda = 1550$ nm, transmit bit-rate 1 Gbps, PD responsibility $\eta$ = 0.9, aperture radius $r_a$ = 5 cm, the link length $Z$ = 20 km, $\alpha = 4$ and $\beta = 1.7$. 

In Figs. \ref{6_1} and \ref{6_2}, we plot the channel distribution for the different values of the optical beam-width $w_z$ at the receiver aperture. We also plot the proposed channel model in \cite{dabiri2018channel} as a benchmark.  According to these two figures, we find that the accuracy of the derived analytical channel model depends on the values of $w_z$ and, for sufficiently large values of $w_z$, a perfect match
between simulations and theory can be realized. For the considered long-range FSO systems, $w_z$ holds the value of several meters, hence, our derived analytical expression for the channel model of such links meets an acceptable degree of accuracy. Accordingly, relevant performance evaluation such as outage probability can be analytically carried out without performing Monte-Carlo simulations. Moreover, from  Figs. \ref{6_1} and \ref{6_2}, one can readily observe that the proposed channel model in \cite{dabiri2018channel}, which is based on Gaussian beam profile for short-range UAV-based link,  is not able to achieve acceptable accuracy for long-range links, i.e., when plane wave regime is assumed for the optical beam at the receiver.

Finally, the curves of outage probability versus $P_t$ under different values of link length in Fig. \ref{outage_div}. As expected, for the same transmit power, $\mathbb{P}_{\textrm{out}}$ increases when the link length is increased. Meanwhile, the results
of this figure verify the accuracy of the derived analytical
expression for the outage probability that makes it easier designing such FSO links without resorting to time-consuming simulations.

\section{CONCLUSION}
\label{con}
A statistical channel model for a long-range ground-to-air FSO link under the plane wave regime was proposed in this paper. Different link parameters, i.e., transmitter divergence angle, receiver's wobbling, jitter due to beam wander, attenuation loss, and atmospheric turbulence were included in the proposed model. Subsequently, a closed-form expression for the outage probability was derived to evaluate the performance of the system. Our analytical results, have
made it possible to study and design such FSO systems without resorting to time-consuming simulations.

\end{document}